\documentclass[aps,prl,twocolumn,noshowpacs,superscriptaddress]{revtex4-1}
\usepackage{graphicx,epsfig, subfigure}
\usepackage{amsbsy}
\usepackage{amssymb}
\usepackage{amsmath}
\usepackage{hyperref}
\usepackage[bottom]{footmisc}
\usepackage{xcolor}
\usepackage{color}

\date{\today}

\begin{document}

\newcommand{\eqnref}[1]{Eq.~\ref{#1}}
\newcommand{\figref}[2][]{Fig.~\ref{#2}#1}

\title{Experimental test of the differential fluctuation theorem and a generalized Jarzynski equality for arbitrary initial states}

\author{Thai M. Hoang}
\thanks{Current address: Sandia National Laboratories, Albuquerque, NM 87123}
 \affiliation{Department of Physics and Astronomy, Purdue University, West Lafayette, IN 47907, USA}%Lines break automatically or can be forced with

 \author{Rui Pan}
 \affiliation{School of Physics, Peking University, Beijing 100871, China}

\author{Jonghoon Ahn}
 \affiliation{School of Electrical and Computer Engineering, Purdue University, West Lafayette, IN 47907, USA}

\author{Jaehoon Bang}
 \affiliation{School of Electrical and Computer Engineering, Purdue University, West Lafayette, IN 47907, USA}

\author{H. T. Quan}
\email{Corresponding author: htquan@pku.edu.cn}
\affiliation{School of Physics, Peking University, Beijing 100871, China}
\affiliation{Collaborative Innovation Center of Quantum Matter, Beijing 100871, China}

\author{Tongcang Li}
 \email{Corresponding author: tcli@purdue.edu}
 \affiliation{Department of Physics and Astronomy, Purdue University, West Lafayette, IN 47907, USA}
 \affiliation{School of Electrical and Computer Engineering, Purdue University, West Lafayette, IN 47907, USA}
 \affiliation{Purdue Quantum Center, Purdue University, West Lafayette, IN 47907, USA}
 \affiliation{Birck Nanotechnology Center, Purdue University, West Lafayette, IN 47907, USA}

\begin{abstract}
{Nonequilibrium processes of small systems such as molecular machines are ubiquitous in biology, chemistry and physics, but are often challenging to comprehend. In the past two decades, several exact thermodynamic relations of nonequilibrium processes, collectively known as fluctuation theorems, have been discovered and provided critical insights. These fluctuation theorems are generalizations of the second law, and can be unified by a differential fluctuation theorem. Here we perform the first experimental test of the differential fluctuation theorem, using an optically levitated nanosphere in both underdamped and overdamped regimes, and in both spatial and velocity spaces. We also test several theorems that can be obtained from it directly, including a generalized Jarzynski equality that is valid for arbitrary initial states, and the Hummer-Szabo relation. %the Crooks fluctuation theorem, and the Jarzynski equality.
 Our study experimentally verifies these fundamental theorems, and initiates the experimental study of stochastic energetics with the instantaneous velocity measurement.
}
\end{abstract}

\maketitle

In the past two decades, there were significant developments in nonequilibrium statistical mechanics of small systems in which thermal fluctuation is influential \cite{jarzynski2011review}.
The most prominent progresses are the discoveries of various fluctuation theorems (FT), which connect  microscopic dynamics with thermodynamic behaviors \cite{jarzynski2011review}. These FT, such as the Jarzynski equality (JE) \cite{jarzynski1997nonequilibrium, liphardt2002equilibrium} and the Crooks fluctuation theorem (CFT) \cite{crooks1999entropy, collin2005verification}, reformulate the inequality of the second law  into equalities, and reveal the universal laws that the fluctuating thermodynamic variables must obey in processes arbitrarily far from thermal equilibrium. As they are refinements of the second law on individual trajectories,
they provide critical understandings of  behaviors of biological systems at the single molecular level \cite{hummer2001free,liphardt2002equilibrium, collin2005verification, HarrisExperimental2007, gupta2011experimental,junier2009recovery,alemany2012experimental} and nonequilibrium dynamics of a wide range of physical systems
\cite{wang2002experimental, trepagnier2004experimental, Douarche2006, bechhoefer2014landauer, GieselerDynamic2014, lee2015nonequilibrium,pekola2015review,seifert2005entropy,kawai2007dissipation, maragakis2008differential, esposito2010detailed,sagawa2010generalized, toyabe2010experimental,gong2015jarzynski,blickle2006,martinez2016carnot,berut2013,Koski2014demon}. While JE and CFT are valid for processes far from thermal equilibrium, 
they require the initial state to be in a thermal equilibrium state.

In 2000, a differential fluctuation theorem (DFT) connecting the joint probabilities of entropy production  and arbitrary generalized coordinates (e.g. position and velocity
coordinates) was derived by Jarzynski \cite{jarzynski2000hamiltonian}. An equivalent DFT for work was derived by Nobel laureate Karplus and his colleagues Maragakis and Spichty in 2008 \cite{maragakis2008differential}. It is remarkable that the DFT can unify various FT as long as detailed balance is not violated \cite{maragakis2008differential}(see Fig. S1 in the supplemental material for the relation between different fluctuation theorems \cite{supportDi}).  The DFT also leads to a generalized Jarzynski equality (GJE) for arbitrary initial states \cite{gong2015jarzynski}. Such ability is rooted from the fact that  most FT originate from the same fundamental principle:  the microscopic reversibility connecting  forward and reverse trajectories \cite{jarzynski2011review,crooks1999thesis, seifert2005entropy, broeck2014review, gong2015jarzynski, jarzynski2000hamiltonian, seifert2012stochastic}.
Testing the DFT and other FT would deepen our understanding of the second law and nonequilibrium physics, including dissipation \cite{kawai2007dissipation}, hysterisis \cite{crooks2008arrow}, and irreversibility \cite{Gomez-Marin2008}. In order to test the DFT at its desired level of detail, we need  large statistics and the ability to  track individual  trajectories of a stochastic process in the phase space \cite{maragakis2008differential}, which requires the measurement of instantaneous velocities of Brownian motion \cite{li2010measurement}.

%\pagebreak
\begin{figure}[t!]
	\includegraphics[scale=1.0]{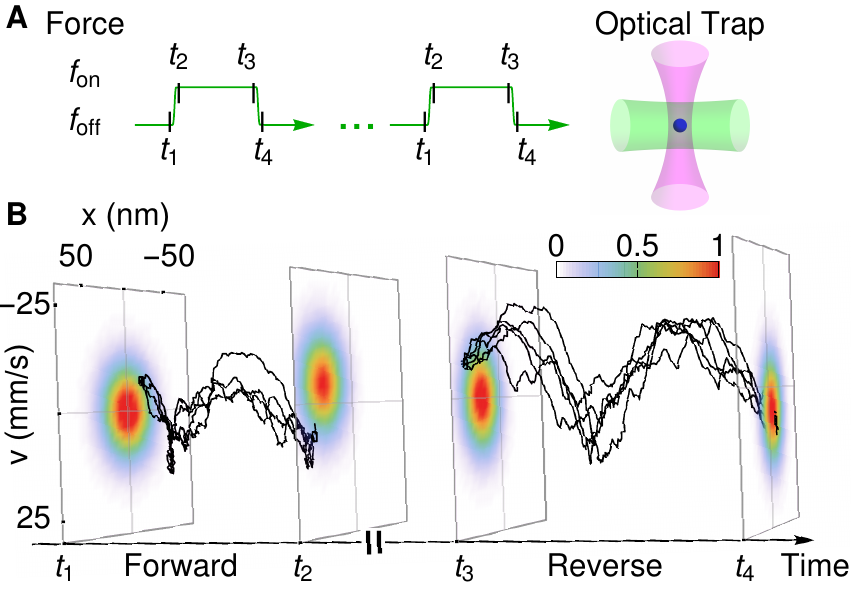}
	\caption{\textbf{A}, Experimental scheme. A silica nanosphere (blue sphere) is trapped in an optical tweezer formed by a  focused 1550-nm laser beam (magenta). A series of 532-nm laser pulses (green) exerts an optical force on the nanosphere to drive nonequilibrium processes. Within each pulse, an optical force is rapidly ramped from $f_\mathrm{off}$ at time $t_1$ to $f_\mathrm{on}$ at time $t_2$ during the forward process (green pulse). The reverse process from time $t_3$ to $t_4$ is the  time-reversed correspondence of the forward process.
	\textbf{B}, An example of experimental data. Vertical slides represent the measured time snapshots of the probability distributions at times $t_1,~t_2,~t_3$, and $t_4$ as illustrated in \textbf{A}.
	Black curves represent experimental phase-space trajectories  during forward processes initialized at   $(x_1, v_1)$ and finalized at $(x_2, v_2)$, and during reverse processes initialized at the  $(x_2, -v_2)$ and finalized at $(x_1, -v_1)$. Here $x_1 =-19$~nm,  $x_2 =55$ ~nm, $v_{1}=-7$~mm/s, and $v_{2}=7$~mm/s.  The nanosphere is levitated in air at 50~torr, and $f_\mathrm{off}=0$, $f_\mathrm{on}=340$~fN.
	}
	\label{Fig:Apparatus}
\end{figure}

In this work, we experimentally test the differential fluctuation theorem \cite{jarzynski2000hamiltonian,maragakis2008differential} using an optically levitated nanosphere which can be  trapped in air continuously for weeks for acquiring large sets of data.  Our ultrasensitive  optical tweezer can measure both instantaneous position and instantaneous  velocity \cite{li2010measurement} of a levitated nanosphere
 to test DFT.
Over one million experimental cycles per setting ($\sim 10^{10}$ position data points per setting with a 10 MHz acquisition rate) provide sufficient statistics to validate the DFT at its desired level of detail, e.g., testing DFT for nonequilibrium processes connecting two  points in the position-velocity  space \cite{maragakis2008differential}. Several fluctuation theorems, including the JE  \cite{jarzynski1997nonequilibrium, liphardt2002equilibrium}, the CFT  \cite{crooks1999entropy, collin2005verification}, the Hummer-Szabo relation (HSR) \cite{hummer2001free, gupta2011experimental, HarrisExperimental2007},  the GJE \cite{gong2015jarzynski, crooks1999thesis, kawai2007dissipation},  the extended fluctuation relation (EFR)  \cite{junier2009recovery, alemany2012experimental}, and the fluctuation theorem for ligand binding (FTLB) \cite{camunas2017} can be unified by the DFT \cite{crooks1999thesis,seifert2005entropy,broeck2014review,gong2015jarzynski}.
We have also tested several such theorems.
In our experiment, the air pressure can be adjusted to  test these theorems in both underdamped and overdamped regimes.
This study demonstrates a powerful approach applicable in exploring a wide range of nonequilibrium systems \cite{hummer2001free, liphardt2002equilibrium, collin2005verification, wang2002experimental, trepagnier2004experimental, lee2015nonequilibrium, Douarche2006, GieselerDynamic2014, HarrisExperimental2007, gupta2011experimental, junier2009recovery, alemany2012experimental}  since a complete description of the stochastic system includes the information of both position and velocity.

\begin{figure}[t!]
	\includegraphics[scale=0.34]{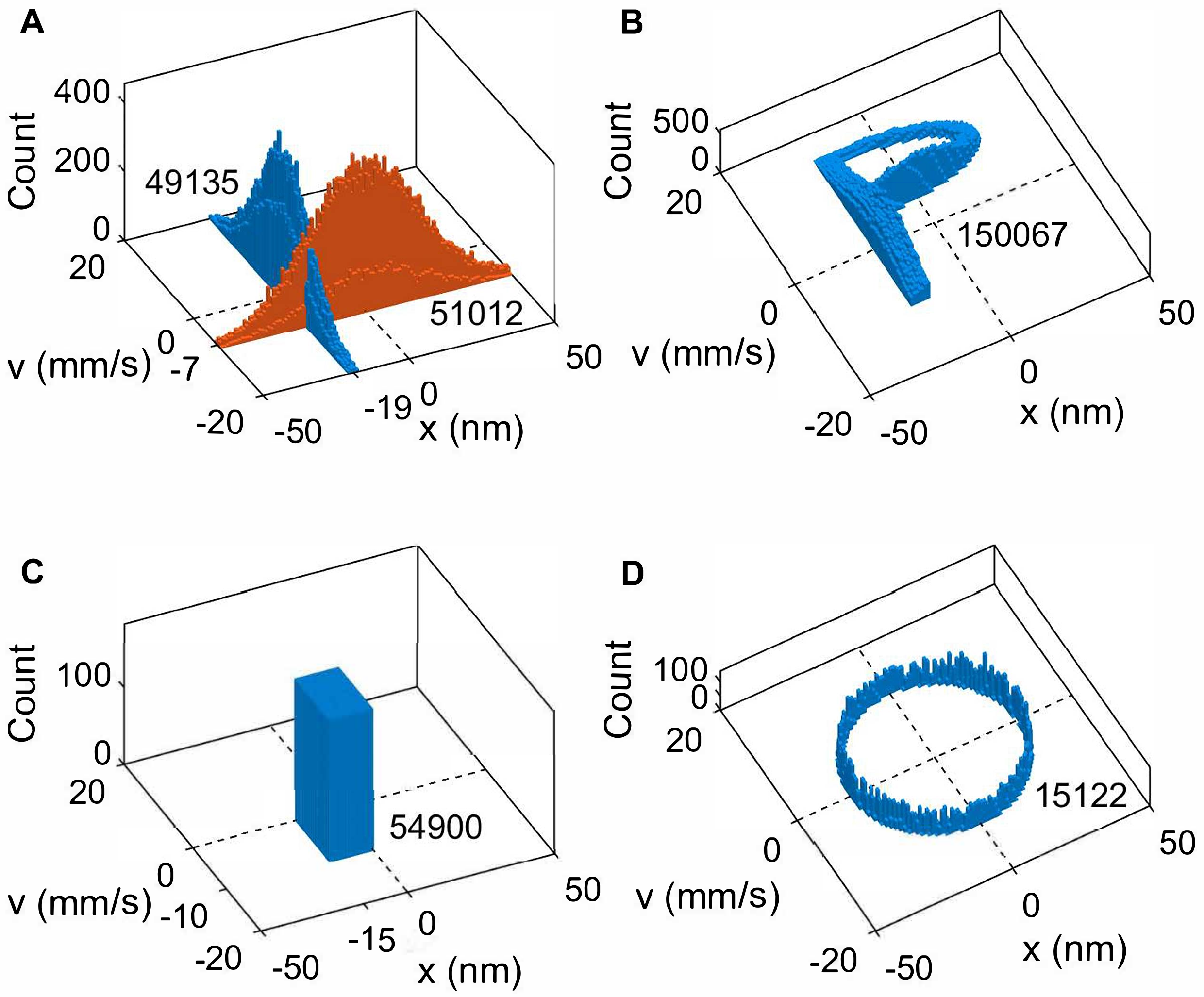}
	\caption{Examples of arbitrary nonequilibrium initial states of trajectory ensembles prepared by an information-based method.  \textbf{A}, Nonequilibrium initial states with narrow distributions in position or velocity.  \textbf{B}, An exotic nonequilibirum state with a P-shaped distribution in phase space. \textbf{C}, Uniform distribution within a rectangle ($- 15$ nm $ < x < 0$, $-10$ mm/s $< v <0$) in the phase space .  \textbf{D}, A microcanonical ensemble with the energy shell $1.3 k_B T< E < 1.35 k_B T.$  The number of experimental trajectories started from each nonequilibrium initial state is labeled next to its distribution.
	}\label{Fig:phasespace}
\end{figure}

Our experiments are carried out using a silica nanosphere levitated by a 1550 nm optical tweezer (\figref{Fig:Apparatus}{A}) \cite{HoangElectron2016}. The nonequilibrium processes are controlled by a force parameter $f$ which is an optical force exerted on the nanosphere by a 532-nm laser beam.
In a forward process, the optical force is ramped from $f_\mathrm{off}$ at time $t_1$  to $f_\mathrm{on}$ at time $t_2$. The reverse process is from $t_3$ to $t_4$.
The DFT connects the forward and reverse processes as \cite{jarzynski2000hamiltonian,maragakis2008differential}:
\begin{eqnarray}
P_R(-W, b^* \rightarrow a^*) \big/ P_F(W, a\rightarrow b) = e^{-\beta(W-\Delta F)},
	\label{EqnDifferential}
\end{eqnarray}
where $a$, $b$ can be arbitrary generalized coordinates.
In our work,  $a$ and $b$ denote the position ($x$)  and/or velocity ($v$) coordinate of a levitated nanosphere,  e.g., $a$ can be  $x$, or $v$, or $(x,v)$.
$P_F(W, a \rightarrow b)$ is the forward joint probability of performing nonequilibrium work $W$ for those trajectories starting from $a$ and ending at $b$, and $P_R(-W, b^* \rightarrow a^*)$ is the reverse joint probability. The work distribution $P_F(W)$ can be obtained by integrating $P_F(W, a \rightarrow b)$ over $a$ and $b$. The asterisk (*) denotes a reversal of the velocity components of $a$ or $b$. $\Delta F = -(f^2_\mathrm{on}- f^2_\mathrm{off})/(2k)$ is the free energy difference between the equilibrium states of the optical forces $f_\mathrm{on}$ and $f_\mathrm{off}$. Here $k$ is the trap stiffness. $\beta=1/(k_B T)$, where $k_B$ is the Boltzmann constant, and $T=296$~K is the room temperature.

%\pagebreak
\begin{figure*}[t!]
	\includegraphics[scale=0.9]{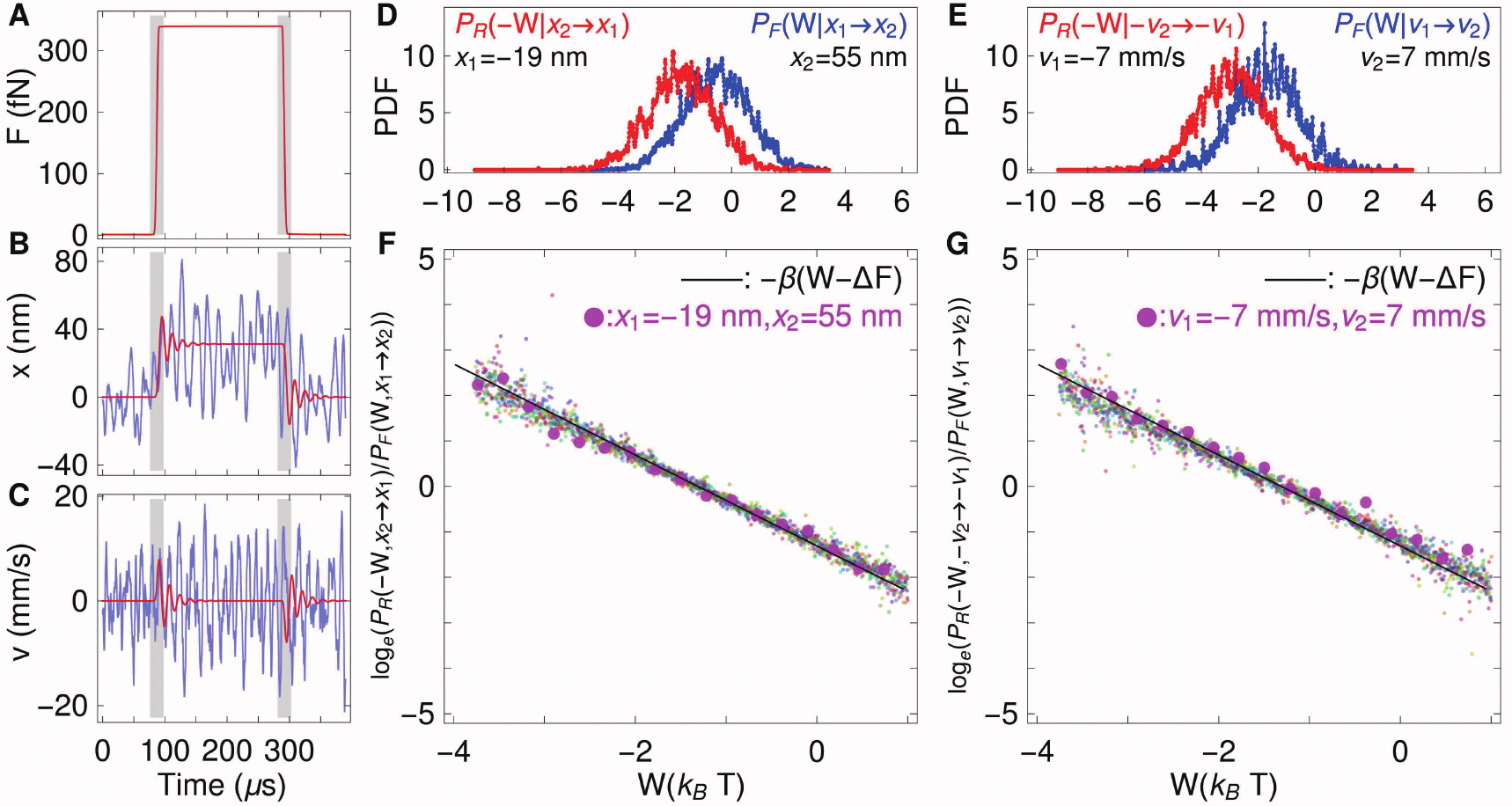}
	\caption{Testing the differential fluctuation theorem in the underdamped regime. \textbf{A}, Optical force.  \textbf{B-C}, Measured position and velocity trajectories. A single trajectory is shown in blue, and the  averaged trajectory of over one million trajectories is shown in red. The gray shaded regions in \textbf{A-C} denote the forward and reverse intervals, respectively. It takes roughly $4.6~\mu \mathrm{s}$ for the optical force strength to switch from 10\% to 90\% level.
	\textbf{D}, An example of probabilities $ P_{F}( W|{x_1\rightarrow x_2})$ and $P_{R}(-W|{x_2\rightarrow x_1} )$ in position coordinate.
	\textbf{E},  An example of probabilities $ P_{F}(W|{v_1\rightarrow v_2})$ and $P_{R}(-W|{-v_2\rightarrow -v_1}) $ in velocity coordinate. The label of horizontal axis in \textbf{D} and \textbf{E} is $W(k_{B}T)$.
	\textbf{F, G} Testing the DFT in position and velocity spaces.  The small markers with different colors represent measurements of $\log_e \frac{P_R(-W, x_2 \rightarrow x_1)}{P_F(W, x_1 \rightarrow x_2)}$ and $\log_e \frac{P_R(-W, -v_2 \rightarrow -v_1)}{P_F(W, v_1 \rightarrow v_2)}$ for 121 different $\{x_1, x_2\}$ and $\{v_1, v_2\}$ combinations, respectively. The big magenta markers are results for parameters shown in \textbf{D} and \textbf{E}, respectively.
	The black lines represent $-\beta(W-\Delta F)$.	
	}
	\label{Fig:DiffUnderdamp}
\end{figure*}

To test the DFT in detail, over one million experimental forward-reverse cycles (500 $\mu$s/cycle) are performed for a given irreversible setting. Their distributions in the position-velocity space are shown in  \figref{Fig:Apparatus}{B}. The driving optical forces significantly shift the distributions away from the undriven ones. Black curves in \figref{Fig:Apparatus}{B} are a few examples of measured trajectories evolving from a given point to a different point in the phase space during forward (or reverse) processes. Due to thermal fluctuation, it is not possible to have two trajectories starting from exactly the same point in the phase space. Here we use $(x,v)$ to represent points within $(x\pm \frac{\sigma_x}{11}, v\pm \frac{\sigma_v}{11})$, where $\sigma_x$ and $\sigma_v$ are the standard deviations of the position and velocity distributions, respectively.

Based on our large sets of experimental data and our ability to measure both the instantaneous velocity and position of a nanoparticle, we develop an efficient method to prepare arbitrary nonequilibrium initial states by conditionally selecting trajectories that start from the desired initial states. Some examples of arbitrary nonequilibrium initial states prepared by our information-based method are shown in \figref{Fig:phasespace}. They are used to test the DFT and the GJE for arbitrary initial states.

\figref{Fig:DiffUnderdamp} shows our experimental results of testing DFT with a 209-nm-radius nanopshere  in the underdamped regime (see supplemental online material for more information \cite{supportDi}).
 The optical force (\figref{Fig:DiffUnderdamp}{A}) is monitored using a fraction of the 532-nm laser   split from the main beam.
\figref{Fig:DiffUnderdamp}{B-C} show the dynamic evolution of the nanosphere in the position and velocity coordinates, respectively. Since the irreversible ramps ($\sim 4.6~\mu$s) are faster than the velocity ($\sim 8.6~\mu\mathrm{s}$) and position ($\sim 100~\mu$s) relaxation times, the nanosphere is far from thermal equilibrium when the ramps finish.

With the acquired position, velocity and force data, the DFT is ready to be tested. The DFT in  \eqnref{EqnDifferential} can be rewritten in the position coordinate as,
 $\frac{P_R(-W, x_2\rightarrow x_1)}{P_F(W, x_1\rightarrow x_2)} = \frac{P_{R}(x_2\rightarrow x_1)}{P_{F}(x_1\rightarrow x_2)} \frac{P_R(-W|{x_2\rightarrow x_1})}{P_F(W|{x_1\rightarrow x_2})}$ \cite{maragakis2008differential}. Here ${P_{F}(x_1\rightarrow x_2)}$ is the probability of having a forward trajectory going from  $x_1$ to $x_2$, and ${P_{R}(x_2\rightarrow x_1)}$ is the probability of a reverse trajectory going from $x_2$ to $x_1$. These quantities can be calculated using the distributions illustrated in the \figref{Fig:Apparatus}{B}. They are essentially equivalent to the number of forward (reverse) trajectories going from  $x_1$ to $x_2$ ( $x_2$ to $x_1$).  ${P_F(W|{x_1\rightarrow x_2})}$ is the probability of performing work $W$ for those forward trajectories going from $x_1$ to $x_2$, and ${P_R(-W|{x_2\rightarrow x_1})}$ is the reverse probability (\figref{Fig:DiffUnderdamp}{D}). Similarly, \figref{Fig:DiffUnderdamp}{E} shows examples of ${P_F(W|{v_1\rightarrow v_2})}$ and ${P_R(-W|{-v_2\rightarrow -v_1})}$ in the velocity coordinate. The minus sign ($-$) in the velocity space  is due to the time reversal symmetry of the reverse process. Here irreversible work is calculated as $W =- \sum_{i=1}^{n-1} (f_{i+1} - f_i) (x_i+x_{i+1})/2$ for $n$ successive position and force measurements. This formula is obtained using the Hamiltonian, $H = \frac{1}{2} kx^2 - f x + \frac{1}{2}mv^2$, and the work definition
$W = \int_{0}^\tau dt \dot{f}(t) \frac{\partial H}{\partial f}$  during a ramp period $\tau$ \cite{jarzynski1997nonequilibrium}. %Here $k$ is the trap stiffness.

%\pagebreak
\begin{figure}[t!]
	\includegraphics[scale=1.0]{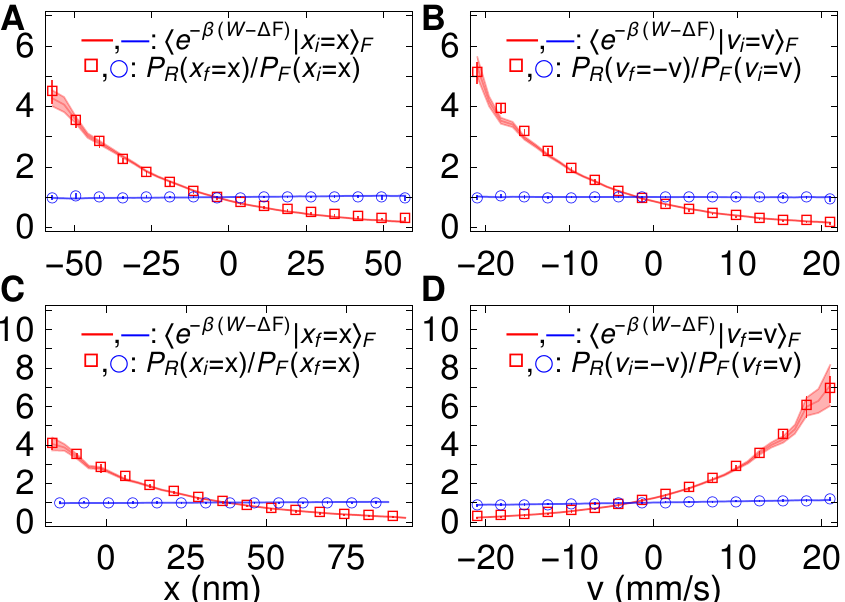}
	\caption{Testing fluctuation theorems in the underdamped regime. \textbf{A, B}, Testing GJE in position and velocity spaces for a fast ramp (red, $4.6~\mu\mathrm{s}$ from 10\% to 90\% levels), and a slow ramp (blue, $40~\mu\mathrm{s}$ from 10\% to 90\% levels). Markers represent the measured $P_{R}(x_f=x)/P_{F}(x_i=x)$ and $P_{R}(v_f=-v)/P_{F}(v_i=v)$ in position and velocity spaces, respectively.
\textbf{C, D}, Testing HSR in position and velocity spaces for a fast ramp (red) and a slow ramp (blue). Markers represent the measured $P_{R}(x_i=x)/P_{F}(x_f=x)$ and $P_{R}(v_i=-v)/P_{F}(v_f=v)$ in position and velocity spaces, respectively.
 The errorbars of $P_R(x)/P_F(x)$ and $P_R(-v)/P_F(v)$ represent the standard deviation of the measurements for 20 equal divisions in each subset $x$ and $v$, respectively. The markers represent their mean values.
 In \textbf{A--D}, the shaded line represent $\langle e^{-\beta(W-\Delta F)} \rangle$, where its thickness represents the uncertainty of 600 work (Joule) calibrations.	
	 }
	\label{Fig:GeneralizedJHS}
\end{figure}

The DFT is tested in detail using 121 different initial-final combinations in the position and velocity coordinates uniformly distributed in $(\pm \sigma_x, \pm \sigma_v)$. \figref{Fig:DiffUnderdamp}{F} shows that the left hand side, $\frac{P_R(-W, x_2\rightarrow x_1)}{P_F(W, x_1\rightarrow x_2)}$, agrees well with the right hand side, $e^{-\beta(W-\Delta F)}$, of the DFT in \eqnref{EqnDifferential}. Here $e^{-\beta(W-\Delta F)}$ is a function of the work variable $W$. The free energy difference can be calculated as $\Delta F = -1.3~k_B T$ with $f_\mathrm{off}/f_\mathrm{on} = 0/340~\mathrm{fN}$  (\figref{Fig:DiffUnderdamp}{A}).
Similarly, we can verify the DFT in the velocity coordinate (\figref{Fig:DiffUnderdamp}{G}). Data points also distribute closely to the curves $e^{-\beta(W-\Delta F)}$.
Thus our data agree with the DFT well in   position and velocity coordinates simultaneously.

Our experimental data can also test other fluctuation theorems which are direct integrations of the DFT \cite{supportDi}. Integrating the  \eqnref{EqnDifferential} over $W$ and $b$, we obtain the GJE for delta initial distributions in the position or velocity coordinates ($a=x$ or $a=v$)    \cite{gong2015jarzynski,crooks1999thesis,kawai2007dissipation},
\begin{eqnarray}
	\langle e^{-\beta(W-\Delta F)}|x_i=x \rangle_{F}   &=& P_{R}(x_f=x) / P_F(x_i=x),  \nonumber\\
	\langle e^{-\beta(W-\Delta F)}|v_i=v \rangle_{F}  &=&
	P_{R}(v_f=-v) / P_F(v_i=v).
	\label{Eqn:GeneralizedJarzynski}
\end{eqnarray}
Here $P_{F}(x_i=x)$ is the probability that a forward trajectory \textit{initializes} at $x$, and $P_{R}(x_f=x)$ is the probability that a reverse trajectory \textit{finalizes} at $x$. We use subscripts ``$i$" and ``$f$" to denote ``initial" and ``final" respectively. Similarly, $P_{F}(v_i=v)$ and $P_{R}(v_f=-v)$ are the probabilities in the velocity coordinate. They are proportional to the number of trajectories \textit{initialized} (\textit{finalized}) at $x$ or $v$ (Fig.~\ref{Fig:phasespace}A).
The value $\langle e^{-\beta(W-\Delta F)}|x_i=x~ (\mathrm{or}~{v_i=v}) \rangle_{F} $ is averaged over all  forward trajectories initialized at $x$ or $v$ in the position or velocity coordinates. The data agree well with the GJE as shown in \figref{Fig:GeneralizedJHS}{A,B}. For slow ramps, the measured  $\langle e^{-\beta(W-\Delta F)}|x_i~ (\mathrm{or}~{v_i}) \rangle_{F}$ stays closely to 1, which is the result of a reversible process. However, for fast (irreversible) ramps, the values of $\langle e^{-\beta(W-\Delta F)}|x_i~ (\mathrm{or}~{v_i}) \rangle_{F}$ diverge away from 1, so the GJE is needed to explain our observations.

Similarly, integrating \eqnref{EqnDifferential} over $W$ and $a$ leads to the HSR \cite{supportDi} in the position and velocity spaces ($b=x$ or $b=v$)  \cite{hummer2001free,kawai2007dissipation, maragakis2008differential},
\begin{eqnarray}
\langle e^{-\beta(W-\Delta F)}|x_f=x \rangle_{F}  &=& P_{R}(x_i=x)/P_{F} (x_f=x), \nonumber \\
\langle e^{-\beta(W-\Delta F)}|v_f=v \rangle_{F} &=& P_{R} (v_i=-v)/P_F(v_f=v).  
\label{Eqn:Hummer-Szabo}
\end{eqnarray}
Here $\langle e^{-\beta(W-\Delta F)}|{x_f=x}~ (\mathrm{or}~ {v_f=v}) \rangle_{F} $, $P_{F}(x_f=x)$, and $P_{R}(x_i=x)$ are denoted using the same conventions as in the GJE in \eqnref{Eqn:GeneralizedJarzynski}. The data agree well with the HSR for both fast (irreversible) ramps and slow (reversible) ramps  as shown in \figref{Fig:GeneralizedJHS}{C,D}.

%The JE \cite{jarzynski1997nonequilibrium} can be obtained by integrating \eqnref{EqnDifferential} over $a,~b$ and $W$. The data show an excellent agreement with JE, $\langle e^{-\beta(W -\Delta F)} \rangle =1.02\pm 0.02$ (\figref{Fig:GeneralizedJHS}{E}).The CFT \cite{crooks1999entropy} can be obtained by integrating \eqnref{EqnDifferential} over $a$ and $b$, $P_R(-W)/P_F(W)  =   e^{-\beta(W-\Delta F)}$. The measurements of $\log_e(P_R(-W)/P_F(W))$ show an excellent agreement with a linear fit (\figref{Fig:GeneralizedJHS}{F}). The fit yields a linear slope of $-0.99 \pm 0.03$ which agrees  well with the theoretical slope of $-1$.

\begin{table}
\begin{ruledtabular}
\begin{tabular}{c|c|c|c|c}
Initial   & Thermal & P-shaped & Uniform & Microcanonical\\
state   & equilibrium & state &distribution & ensemble  \\
         &  & (Fig.~\ref{Fig:phasespace}B) & (Fig.~\ref{Fig:phasespace}C) &(Fig.~\ref{Fig:phasespace}D)  \\
\hline
lhs & 1.02$\pm$0.02 &0.92$\pm$0.02 & 1.42$\pm$0.03 & 1.08$\pm$0.02\\
\hline
rhs &1  &0.90 &1.42 & 1.07 \\
\end{tabular}
\end{ruledtabular}
\caption{Test of the GJE for arbitrary initial states. The thermal equilibrium  state and three representative nonequilibrium initial states shown in Fig.~\ref{Fig:phasespace}B, \ref{Fig:phasespace}C, \ref{Fig:phasespace}D are chosen for the test. The second and third rows show the data of the lhs and rhs of \eqnref{Eqn:JEarbitrary} for each initial state, respectively.  }
\label{tab:1}
\end{table}

Integraging Eq.~\ref{Eqn:GeneralizedJarzynski} over initial phase space points with an arbitrary initial distribution, one obtains the GJE for arbitrary initial states proposed in Ref. \cite{gong2015jarzynski}
\begin{equation}
\begin{split}
&\left\langle e^{-\beta(W-\Delta F)}\right\rangle_{P_{ini}(x_i,v_i)} =	\label{Eqn:JEarbitrary} \\
 &\int \frac{P_{R}(x_f=x,v_f=-v)}{P_{F}(x_i=x,v_i=v)} P_{ini}(x_i=x,v_i=v) dxdv,
\end{split}
\end{equation}
where $P_{ini}(x_i,v_i)$ indicates an arbitrary initial distribution in phase space. We test Eq.~\ref{Eqn:JEarbitrary} for the thermal equilibrium initial state and three representative nonequilibrium initial states as shown in Fig.~\ref{Fig:phasespace}B, \ref{Fig:phasespace}C, \ref{Fig:phasespace}D. The results are shown in TABLE \ref{tab:1}. The left hand side (lhs) and right hand side (rhs) of \eqnref{Eqn:JEarbitrary} agree well with each other within the experimental uncertainty.

With our experimental data we can test JE \cite{jarzynski1997nonequilibrium} and CFT \cite{crooks1999entropy} with high precision. The results are shown in Fig. S2 in the supplemental material \cite{supportDi}. For completeness, we also tested the DFT in the overdamped regime  ($a=x_1$ and $b=x_2$) where  the velocity relaxes to equilibrium much faster than other processes.  As shown in Fig. S3 in the supplemental material \cite{supportDi}, our experimental data show good agreements with %the JE \cite{jarzynski1997nonequilibrium}, the CFT \cite{crooks1999entropy},
HSR \cite{hummer2001free}, the DFT \cite{maragakis2008differential}, and the GJE \cite{kawai2007dissipation,gong2015jarzynski}.
Overall, the differential fluctuation theorem unifies many existing fluctuation theorems \cite{crooks1999thesis,seifert2005entropy,broeck2014review,gong2015jarzynski,junier2009recovery, alemany2012experimental, camunas2017}, such as JE, CFT, HSR, GJE, EFR, and FTLB, and is arguably the most detailed fluctuation theorem that can be tested experimentally. The DFT can also improve free energy calculations \cite{maragakis2008differential}. Our experimental results validate the DFT \cite{maragakis2008differential} well in both underdamped and overdamped regimes.
Our work deepens our understanding of  the second law to an unprecedentedly detailed level.  It initiates the experimental study of stochastic thermodynamics with instantaneous velocity measurements, and may
shed new light on our understanding of the origin of time's arrow \cite{crooks2008arrow}. Once cooled to the quantum regime, a levitated nanosphere in vacuum can be used to investigate quantum nonequilibrium thermodynamics in the mesoscopic regime \cite{yin2013}. This system can also be used to study effects of geometry in thermodynamic control \cite{zulkowski2012geometry}.

\begin{acknowledgments} 
T.L. acknowledges support from NSF under Grant No. PHY-1555035 and the Tellabs Foundation. H.T.Q.  acknowledges support from the National Science Foundation of China under grants 11375012, 11534002, and The Recruitment Program of Global Youth Experts of China.
\end{acknowledgments}

%\section*{Data availability}
%The data that support the findings of this study are available from the corresponding authors upon reasonable request.

%\section*{Author contributions}
%T.L., H.T.Q., T.M.H.,  and R.P. conceived and designed the project.  T.M.H., J.A, J.B. built the experimental apparatus. T.M.H. performed measurements. T.M.H. and R.P. analyzed the data. T.L. and H.T.Q. supervised the work. All authors co-wrote the paper.

%\section*{Competing financial interests}
%The authors declare no competing financial interests.

%\bibliographystyle{unsrt}
%\bibliographystyle{nature}

%

\newpage
\renewcommand{\thefigure}{S\arabic{figure}}
\setcounter{figure}{0}
\onecolumngrid
\section{Supplemental Material}

\subsection{Relation between the differential fluctuation theorem and other fluctuation theorems}

\begin{figure*}[bp]
	\includegraphics[scale=0.7]{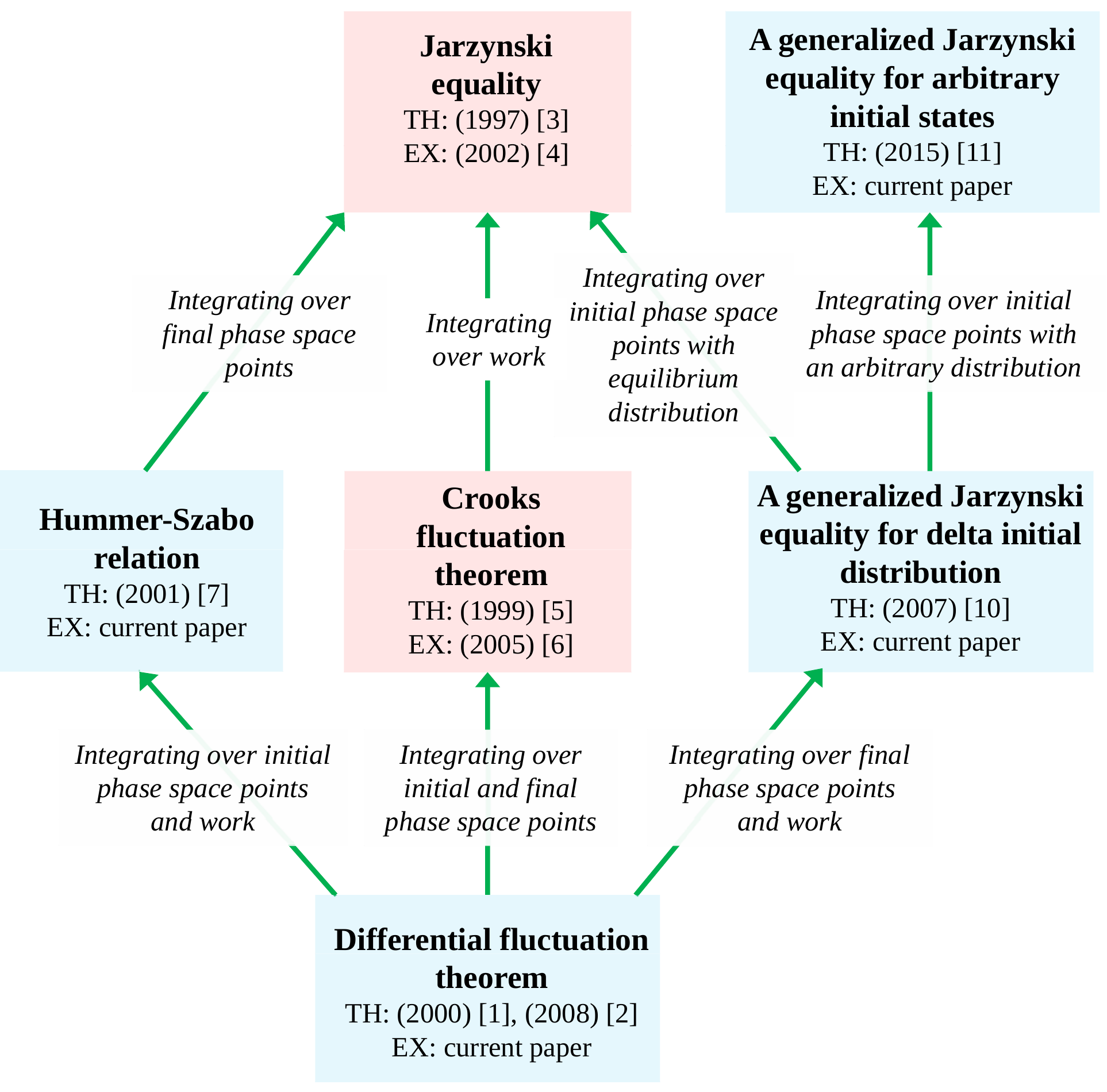}
	\caption{Relation between the DFT and other FT. Here ``TH'' indicates ``theory'', and ``EX'' indicates ``experiment''. The number in a round bracket indicates the year in which the theory was proposed or the experiment was carried out, and the number in a square bracket indicates the reference source. It can be seen that the JE and the CFT have been tested experimentally before, which are highlighted with red background. The DFT, the HSR,  the GJE  for delta initial distribution and GJE  for arbitrary initial states are tested in the current study. They are highlighted with blue background. 
	}
	\label{Fig:Hierarchy}
\end{figure*}

As discussed in the main text, the differential fluctuation theorem (DFT) \cite{supportjarzynski2000hamiltonian,supportmaragakis2008differential} unifies many existing fluctuation theorems, such as  the Jarzynski equality (JE) \cite{supportjarzynski1997nonequilibrium, supportliphardt2002equilibrium}, the Crooks fluctuation theorem (CFT) \cite{supportcrooks1999entropy, supportcollin2005verification}, the Hummer-Szabo relation (HSR) \cite{supporthummer2001free, supportgupta2011experimental, supportHarrisExperimental2007}, the generalized Jarzynski equality (GJE) for the delta initial distribution\cite{supportkawai2007dissipation} and a GJE for arbitrary initial states proposed in Ref.~\cite{supportgong2015jarzynski}.
\figref{Fig:Hierarchy} shows the relation between the differential fluctuation theorem \cite{supportjarzynski2000hamiltonian,supportmaragakis2008differential} and other fluctuation theorems.
 The DFT is arguably the most detailed fluctuation theorem that can be tested experimentally.   It originates from the microscopic reversibility of each trajectory. The CFT can be obtained by integrating the DFT over initial and final phase space points. The HSR can be obtained by integrating the DFT over initial phase space points and work. The GJE for delta initial distribution can be obtained by integrating the DFT over final phase space points and work. The JE can be obtained by integrating the CFT over work, or by integrating the HSR over final phase space points, or by integrating the generalized JE over initial phase space points with equilibrium distribution. We note that the CFT, the HSR, and the GJE can not be obtained directly from each other. The GJE  for arbitrary initial distribution can be obtained by integrating the GJE  for delta initial distribution over initial phase space points with an arbitrary distribution. Among theorems listed in Fig. \ref{Fig:Hierarchy}, the JE and the CFT have been tested experimentally before. While some particular cases of the GJE have been investigated elsewhere \cite{supportkawai2007dissipation,supportsagawa2010generalized,supportcilliberto2013erasure,supportpekola2014information}, we used data of trajectory ensembles with arbitrary nonequilibrium initial states to perform the first experimental test of the generalized Jarzynski equality for arbitrary initial states proposed in \cite{supportgong2015jarzynski}.

\subsection{Calibrations in underdamped regime}

The particle position-velocity ($x$-$v$) is calibrated to SI units using the equipartition theorem, $k\langle x^2\rangle = k_B T$.
$k=(4/3)\pi r^3 \rho \Omega^2$ is the trap stiffness. The silica nanosphere has a mass density of $\rho=1960~\mathrm{kg/m^3}$ (Bangs Laboratories). The power spectra and position autocorrelation function \cite{supportHoangElectron2016}  yield a hydrodynamic radius $r = 209 \pm 9$~nm and a trapping frequency $\Omega / 2\pi = 60.4 \pm 0.3 ~ \mathrm{kHz}$.  The optical force is calibrated using Hooke's law, $f=k\Delta x$, with $\Delta x$ being the position displacement. The work (Joule) calibration is the product of the position and force calibrations.

The force data shown in Fig. 3{A} of the main text indicate that forward and reverse forces have good time-reversal symmetry with only $\sim$1\% difference in the area under the force curve. The particle velocity shown in Fig. 3{C} of the main text is calculated after binning 7 position points together to reduce the detection noise \cite{supportli2010measurement}. The nanosphere velocity has a signal-to-noise ratio of about $8.9$.  Here the signal-to-noise ratio  of the nanosphere velocity  is
\begin{eqnarray}
	v_\mathrm{SNR} = \frac{\Delta v_\mathrm{nanosphere}}{\Delta v_\mathrm{without~nanosphere}}.
\end{eqnarray}
$\Delta v_\mathrm{nanosphere}$ is the standard deviation of the nanosphere velocity (signal), and $\Delta v_\mathrm{without~nanosphere}$ is the standard deviation of the velocity without nanosphere (noise).

\subsection{Test of JE and CFT with a levitated nanoparticle in the underdamped regime}
\begin{figure}[h!]
	\includegraphics[scale=1.0]{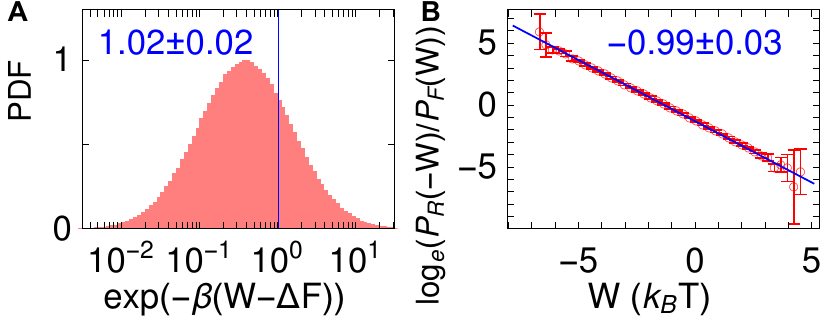}
	\caption{Testing JE and CFT in the underdamped regime.
	\textbf{A}, Testing the JE using forward irreversible work, $\langle e^{-\beta(W -\Delta F)} \rangle =1.02 \pm 0.02$ (vertical line).
	\textbf{B}, Testing the CFT. The slope of the linear fit yields a value $-0.99\pm 0.03$.
	 The errorbars and the uncertainty are due to the standard deviation of 600 work (Joule) calibrations.
	 }
	\label{Fig:JECFT}
\end{figure}

 The JE \cite{supportjarzynski1997nonequilibrium} can be obtained by integrating Eq. 1 in the main text over $a,~b$ and $W$. Our experimental data show an excellent agreement with JE, $\langle e^{-\beta(W -\Delta F)} \rangle =1.02\pm 0.02$ (\figref{Fig:JECFT}{A}). The CFT \cite{supportcrooks1999entropy} can be obtained by integrating Eq. 1 in the main text over $a$ and $b$, $P_R(-W)/P_F(W)  =   e^{-\beta(W-\Delta F)}$. The measurements of $\log_e(P_R(-W)/P_F(W))$ show an excellent agreement with a linear fit (\figref{Fig:JECFT}{B}). The fit yields a linear slope of $-0.99 \pm 0.03$ which agrees  well with the theoretical slope of $-1$.

\subsection{Overdamped regime}

\begin{figure*}[t]
	\includegraphics[scale=0.9]{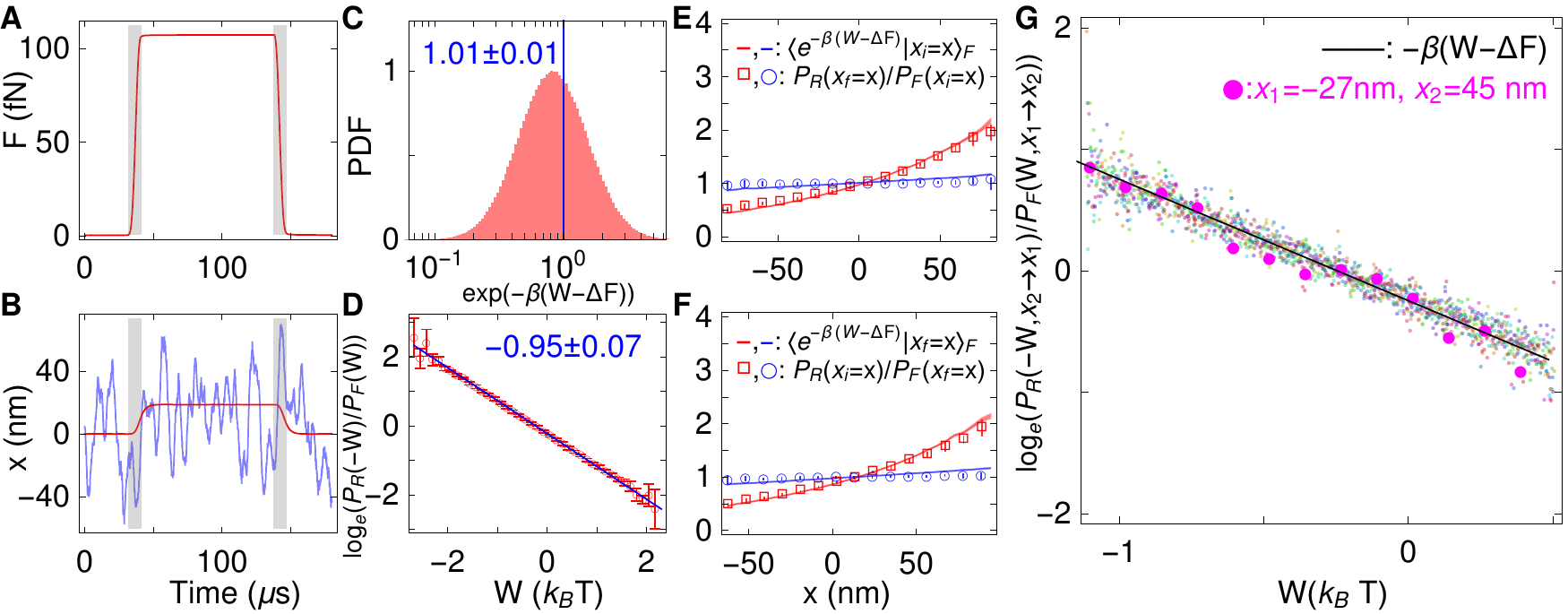}
	\caption{Testing DFT and FT in the overdamped regime. \textbf{A}, The optical force. \textbf{B}, The average of over 1 million position trajectories is shown in red, and a single trajectory is shown in blue.
	\textbf{C}, Testing the JE using the forward irreversible work, $\langle e^{-\beta(W -\Delta F)} \rangle =1.01 \pm 0.01$ (vertical line).
	\textbf{D}, Testing the CFT. The slope of the linear fit yields $-0.95\pm 0.07$.
	In \textbf{C--D}, The errorbars and the uncertainty are the standard deviation of 600 work (Joule) calibrations.
	\textbf{E}, Testing the GJE in position space for a fast ramp (red, $4.8~\mu\mathrm{s}$ from 10\% to 90\% level), and a slow ramp (blue, $62~\mu\mathrm{s}$ from 10\% to 90\% level).
\textbf{F}, Testing the HSR in position space for a fast ramp, and a slow ramp.
In \textbf{E-F}, the shaded line represent $\langle e^{-\beta(W-\Delta F)} \rangle$, where its thickness represents the uncertainty due to 600 work (Joule) calibrations.
The errorbars of $P_{R}(x_i=x)/P_{F}(x_f=x)$ represent the standard deviation of the measurements for 20 equal divisions within the range between $x-\frac{\sigma_x}{11}$ and $x+\frac{\sigma_x}{11}$. The markers represent their mean values.
\textbf{G} Testing differential fluctuation theorem in position space. The small markers with different colors represent the measurements of $\log_e \frac{P_R(-W, x_2 \rightarrow x_1)}{P_F(W, x_1 \rightarrow x_2) }$ for 121 different  combinations $\{x_1, x_2\}$. The large magenta markers illustrate data for the combination $\{-27$~nm, $45$~nm$\}$.
The black line represents $-\beta(W-\Delta F)$.
	}
	\label{Fig:Overdamped}
\end{figure*}

For completeness, we also tested the DFT in the overdamped regime  ($a=x_1$ and $b=x_2$) where  the velocity relaxes to equilibrium much faster than other processes. So it is sufficient to measure the position only. A smaller nanosphere  with a hydrodynamic radius $r = 145\pm 5$~nm is levitated at a pressure of 760 torr. The trapping frequency is $\Omega = 76 \pm 3  ~(2\pi \cdot \mathrm{kHz})$. The optical force profile and the position trajectories are shown in  \figref{Fig:Overdamped}{A,B}. For a typical optical force, $f_\mathrm{off}/f_\mathrm{on}=0/107$~fN, the free energy difference is $\Delta F = -0.24~k_B T$. As shown in \figref{Fig:Overdamped}{B}, the nanosphere position is far from the equilibrium when the ramps finish.
The experimental data show good agreements with the JE \cite{supportjarzynski1997nonequilibrium}, the CFT \cite{supportcrooks1999entropy}, the GJE \cite{supportgong2015jarzynski}, HSR \cite{supporthummer2001free}, and the DFT \cite{supportmaragakis2008differential} as shown in \figref{Fig:Overdamped}{C--G}.

\end{document}